

\catcode`@=11

\def\nofirstpagenoten{\nopagenumbers\footline={\ifnum\pageno>1\tenrm
\hss\folio\hss\fi}}
\def\nofirstpagenotwelve{\nopagenumbers\footline={\ifnum\pageno>1\twelverm
\hss\folio\hss\fi}}
\def\leaderfill{\leaders\hbox to 1em{\hss.\hss}\hfill}
\def\ft#1#2{{\textstyle{{#1}\over{#2}}}}
\def\frac#1/#2{\leavevmode\kern.1em
\raise.5ex\hbox{\the\scriptfont0 #1}\kern-.1em/\kern-.15em
\lower.25ex\hbox{\the\scriptfont0 #2}}
\def\sfrac#1/#2{\leavevmode\kern.1em
\raise.5ex\hbox{\the\scriptscriptfont0 #1}\kern-.1em/\kern-.15em
\lower.25ex\hbox{\the\scriptscriptfont0 #2}}

\parindent=20pt
\def\narrow{\advance\leftskip by 40pt \advance\rightskip by 40pt}

\def\nonarrower{\advance\leftskip by -40pt\advance\rightskip by -40pt}

\def\boxit#1{\vbox{\hrule\hbox{\vrule\kern3pt
        \vbox{\kern3pt#1\kern3pt}\kern3pt\vrule}\hrule}}

\def\gtorder{\mathrel{\raise.3ex\hbox{$>$}\mkern-14mu
             \lower0.6ex\hbox{$\sim$}}}
\def\ltorder{\mathrel{\raise.3ex\hbox{$<$}|mkern-14mu
             \lower0.6ex\hbox{\sim$}}}
\def\dalemb#1#2{{\vbox{\hrule height .#2pt
        \hbox{\vrule width.#2pt height#1pt \kern#1pt
                \vrule width.#2pt}
        \hrule height.#2pt}}}
\def\square{\mathord{\dalemb{4.9}{5}\hbox{\hskip1pt}}}

\font\fourteentt=cmtt10 scaled \magstep2
\font\fourteenbf=cmbx12 scaled \magstep1
\font\fourteenrm=cmr12 scaled \magstep1
\font\fourteeni=cmmi12 scaled \magstep1
\font\fourteenssr=cmss12 scaled \magstep1
\font\fourteenmbi=cmmib10 scaled \magstep2
\font\fourteensy=cmsy10 scaled \magstep2
\font\fourteensl=cmsl12 scaled \magstep1
\font\fourteenex=cmex10 scaled \magstep2
\font\fourteenit=cmti12 scaled \magstep1
\font\twelvett=cmtt12 \font\twelvebf=cmbx12
\font\twelverm=cmr12  \font\twelvei=cmmi12
\font\twelvessr=cmss12 \font\twelvembi=cmmib10 scaled \magstep1
\font\twelvesy=cmsy10 scaled \magstep1
\font\twelvesl=cmsl12 \font\twelveex=cmex10 scaled \magstep1
\font\twelveit=cmti12
\font\tenssr=cmss10 \font\tenmbi=cmmib10
 
 \font\ninebf=cmbx9
\font\ninerm=cmr9  \font\ninei=cmmi9
\font\ninesy=cmsy9 \font\ninessr=cmss9
\font\ninembi=cmmib10 scaled 900
\font\eightit=cmti8 \font\eightsl=cmsl8
\font\eighttt=cmtt8 \font\eightbf=cmbx8
\font\eightrm=cmr8  \font\eighti=cmmi8
\font\eightsy=cmsy8 \font\eightex=cmex10 scaled 800
\font\eightssr=cmss8 \font\eightmbi=cmmib10 scaled 800
 
\font\sevenbf=cmbx7 \font\sevenrm=cmr7 \font\seveni=cmmi7
\font\sevensy=cmsy7 
\font\sevenssr=cmss9 scaled 778 \font\sevenmbi=cmmib10 scaled 700
 
 \font\sixbf=cmbx7 scaled 875
\font\sixrm=cmr6  \font\sixi=cmmi6
\font\sixsy=cmsy6 \font\sixssr=cmss8 scaled 750
\font\sixmbi=cmmib10 scaled 600
\font\fivessr=cmss8 scaled 625  \font\fivembi=cmmib10 scaled 500

\newskip\ttglue
\newfam\ssrfam
\newfam\mbifam

\mathchardef\alpha="710B
\mathchardef\beta="710C
\mathchardef\gamma="710D
\mathchardef\delta="710E
\mathchardef\epsilon="710F
\mathchardef\zeta="7110
\mathchardef\eta="7111
\mathchardef\theta="7112
\mathchardef\iota="7113
\mathchardef\kappa="7114
\mathchardef\lambda="7115
\mathchardef\mu="7116
\mathchardef\nu="7117
\mathchardef\xi="7118
\mathchardef\pi="7119
\mathchardef\rho="711A
\mathchardef\sigma="711B
\mathchardef\tau="711C
\mathchardef\upsilon="711D
\mathchardef\phi="711E
\mathchardef\chi="711F
\mathchardef\psi="7120
\mathchardef\omega="7121
\mathchardef\varepsilon="7122
\mathchardef\vartheta="7123
\mathchardef\varpi="7124
\mathchardef\varrho="7125
\mathchardef\varsigma="7126
\mathchardef\varphi="7127
\mathchardef\partial="7140

\def\fourteenpoint{\def\rm{\fam0\fourteenrm}
\textfont0=\fourteenrm \scriptfont0=\tenrm \scriptscriptfont0=\sevenrm
\textfont1=\fourteeni \scriptfont1=\teni \scriptscriptfont1=\seveni
\textfont2=\fourteensy \scriptfont2=\tensy \scriptscriptfont2=\sevensy
\textfont3=\fourteenex \scriptfont3=\fourteenex \scriptscriptfont3=\fourteenex
\def\it{\fam\itfam\fourteenit} \textfont\itfam=\fourteenit
\def\sl{\fam\slfam\fourteensl} \textfont\slfam=\fourteensl
\def\bf{\fam\bffam\fourteenbf} \textfont\bffam=\fourteenbf
\scriptfont\bffam=\tenbf \scriptscriptfont\bffam=\sevenbf
\def\tt{\fam\ttfam\fourteentt} \textfont\ttfam=\fourteentt
\def\ssr{\fam\ssrfam\fourteenssr} \textfont\ssrfam=\fourteenssr
\scriptfont\ssrfam=\tenmbi \scriptscriptfont\ssrfam=\sevenmbi
\def\mbi{\fam\mbifam\fourteenmbi} \textfont\mbifam=\fourteenmbi
\scriptfont\mbifam=\tenmbi \scriptscriptfont\mbifam=\sevenmbi
\tt \ttglue=.5em plus .25em minus .15em
\normalbaselineskip=16pt
\bigskipamount=16pt plus5pt minus5pt
\medskipamount=8pt plus3pt minus3pt
\smallskipamount=4pt plus1pt minus1pt
\abovedisplayskip=16pt plus 4pt minus 12pt
\belowdisplayskip=16pt plus 4pt minus 12pt
\abovedisplayshortskip=0pt plus 4pt
\belowdisplayshortskip=9pt plus 4pt minus 6pt
\parskip=5pt plus 1.5pt
\twelvefoot
\setbox\strutbox=\hbox{\vrule height12pt depth5pt width0pt}
\let\sc=\tenrm
\let\big=\fourteenbig \normalbaselines\rm}
\def\fourteenbig#1{{\hbox{$\left#1\vbox to12pt{}\right.\n@space$}}
\def\square{\mathord{\dalemb{6.8}{7}\hbox{\hskip1pt}}}}

\def\twelvepoint{\def\rm{\fam0\twelverm}
\textfont0=\twelverm \scriptfont0=\ninerm \scriptscriptfont0=\sevenrm
\textfont1=\twelvei \scriptfont1=\ninei \scriptscriptfont1=\seveni
\textfont2=\twelvesy \scriptfont2=\ninesy \scriptscriptfont2=\sevensy
\textfont3=\twelveex \scriptfont3=\twelveex \scriptscriptfont3=\twelveex
\def\it{\fam\itfam\twelveit} \textfont\itfam=\twelveit
\def\sl{\fam\slfam\twelvesl} \textfont\slfam=\twelvesl
\def\bf{\fam\bffam\twelvebf} \textfont\bffam=\twelvebf
\scriptfont\bffam=\ninebf \scriptscriptfont\bffam=\sevenbf
\def\tt{\fam\ttfam\twelvett} \textfont\ttfam=\twelvett
\def\ssr{\fam\ssrfam\twelvessr} \textfont\ssrfam=\twelvessr
\scriptfont\ssrfam=\ninessr \scriptscriptfont\ssrfam=\sevenssr
\def\mbi{\fam\mbifam\twelvembi} \textfont\mbifam=\twelvembi
\scriptfont\mbifam=\ninembi \scriptscriptfont\mbifam=\sevenmbi
\tt \ttglue=.5em plus .25em minus .15em
\normalbaselineskip=14pt
\bigskipamount=14pt plus4pt minus4pt
\medskipamount=7pt plus2pt minus2pt
\abovedisplayskip=14pt plus 3pt minus 10pt
\belowdisplayskip=14pt plus 3pt minus 10pt
\abovedisplayshortskip=0pt plus 3pt
\belowdisplayshortskip=8pt plus 3pt minus 5pt
\parskip=3pt plus 1.5pt
\tenfoot
\setbox\strutbox=\hbox{\vrule height10pt depth4pt width0pt}
\let\sc=\ninerm
\let\big=\twelvebig \normalbaselines\rm}
\def\twelvebig#1{{\hbox{$\left#1\vbox to10pt{}\right.\n@space$}}
\def\square{\mathord{\dalemb{5.9}{6}\hbox{\hskip1pt}}}}

\def\tenpoint{\def\rm{\fam0\tenrm}
\textfont0=\tenrm \scriptfont0=\sevenrm \scriptscriptfont0=\fiverm
\textfont1=\teni \scriptfont1=\seveni \scriptscriptfont1=\fivei
\textfont2=\tensy \scriptfont2=\sevensy \scriptscriptfont2=\fivesy
\textfont3=\tenex \scriptfont3=\tenex \scriptscriptfont3=\tenex
\def\it{\fam\itfam\tenit} \textfont\itfam=\tenit
\def\sl{\fam\slfam\tensl} \textfont\slfam=\tensl
\def\bf{\fam\bffam\tenbf} \textfont\bffam=\tenbf
\scriptfont\bffam=\sevenbf \scriptscriptfont\bffam=\fivebf
\def\tt{\fam\ttfam\tentt} \textfont\ttfam=\tentt
\def\ssr{\fam\ssrfam\tenssr} \textfont\ssrfam=\tenssr
\scriptfont\ssrfam=\sevenssr \scriptscriptfont\ssrfam=\fivessr
\def\mbi{\fam\mbifam\tenmbi} \textfont\mbifam=\tenmbi
\scriptfont\mbifam=\sevenmbi \scriptscriptfont\mbifam=\fivembi
\tt \ttglue=.5em plus .25em minus .15em
\normalbaselineskip=12pt
\bigskipamount=12pt plus4pt minus4pt
\medskipamount=6pt plus2pt minus2pt
\abovedisplayskip=12pt plus 3pt minus 9pt
\belowdisplayskip=12pt plus 3pt minus 9pt
\abovedisplayshortskip=0pt plus 3pt
\belowdisplayshortskip=7pt plus 3pt minus 4pt
\parskip=0.0pt plus 1.0pt
\eightfoot
\setbox\strutbox=\hbox{\vrule height8.5pt depth3.5pt width0pt}
\let\sc=\eightrm
\let\big=\tenbig \normalbaselines\rm}
\def\tenbig#1{{\hbox{$\left#1\vbox to8.5pt{}\right.\n@space$}}
\def\square{\mathord{\dalemb{4.9}{5}\hbox{\hskip1pt}}}}

\def\eightpoint{\def\rm{\fam0\eightrm}
\textfont0=\eightrm \scriptfont0=\sixrm \scriptscriptfont0=\fiverm
\textfont1=\eighti \scriptfont1=\sixi \scriptscriptfont1=\fivei
\textfont2=\eightsy \scriptfont2=\sixsy \scriptscriptfont2=\fivesy
\textfont3=\eightex \scriptfont3=\eightex \scriptscriptfont3=\eightex
\def\it{\fam\itfam\eightit} \textfont\itfam=\eightit
\def\sl{\fam\slfam\eightsl} \textfont\slfam=\eightsl
\def\bf{\fam\bffam\eightbf} \textfont\bffam=\eightbf
\scriptfont\bffam=\sixbf \scriptscriptfont\bffam=\fivebf
\def\tt{\fam\ttfam\eighttt} \textfont\ttfam=\eighttt
\def\ssr{\fam\ssrfam\eightssr} \textfont\ssrfam=\eightssr
\scriptfont\ssrfam=\sixssr \scriptscriptfont\ssrfam=\fivessr
\def\mbi{\fam\mbifam\eightmbi} \textfont\mbifam=\eightmbi
\scriptfont\mbifam=\sixmbi \scriptscriptfont\mbifam=\fivembi
\tt \ttglue=.5em plus .25em minus .15em
\normalbaselineskip=9pt
\bigskipamount=9pt plus3pt minus3pt
\medskipamount=5pt plus2pt minus2pt
\abovedisplayskip=9pt plus 3pt minus 9pt
\belowdisplayskip=9pt plus 3pt minus 9pt
\abovedisplayshortskip=0pt plus 3pt
\belowdisplayshortskip=5pt plus 3pt minus 4pt
\parskip=0.0pt plus 1.0pt
\setbox\strutbox=\hbox{\vrule height8.5pt depth3.5pt width0pt}
\let\sc=\sixrm
\let\big=\eightbig \normalbaselines\rm}
\def\eightbig#1{{\hbox{$\left#1\vbox to6.5pt{}\right.\n@space$}}
\def\square{\mathord{\dalemb{3.9}{4}\hbox{\hskip1pt}}}}

\def\vfootnote#1{\insert\footins\bgroup\footsuite
    \interlinepenalty=\interfootnotelinepenalty
    \splittopskip=\ht\strutbox
    \splitmaxdepth=\dp\strutbox \floatingpenalty=20000
    \leftskip=0pt \rightskip=0pt \spaceskip=0pt \xspaceskip=0pt
    \textindent{#1}\footstrut\futurelet\next\fo@t}
\def\hangfootnote#1{\edef\@sf{\spacefactor\the\spacefactor}#1\@sf
    \insert\footins\bgroup\footsuite
    \let\par=\endgraf
    \interlinepenalty=\interfootnotelinepenalty
    \splittopskip=\ht\strutbox
    \splitmaxdepth=\dp\strutbox \floatingpenalty=20000
    \leftskip=0pt \rightskip=0pt \spaceskip=0pt \xspaceskip=0pt
    \smallskip\item{#1}\bgroup\strut\aftergroup\@foot\let\next}
\def\footsuite{}
\def\twelvefoot{\def\footsuite{\twelvepoint}}
\def\tenfoot{\def\footsuite{\tenpoint}}
\def\eightfoot{\def\footsuite{\eightpoint}}
\catcode`@=12

\magnification=1200

\rightline{UG-4/92}
\rightline{CTP TAMU-46/92}
\rightline{June 1992}
\vskip 2truecm
\centerline{\bf  Self-Dual Supergravity Theories in $2+2$ Dimensions}
\vskip 2truecm
\centerline{{\bf Eric BERGSHOEFF} }
\vskip .5truecm
\centerline{Institute for Theoretical Physics}
\centerline{University of Groningen}
\centerline{Nijenborgh 4, 9747 AG Groningen}
\centerline{The Netherlands}
\vskip .5truecm
\centerline{and}
\vskip .5truecm
\centerline{{\bf Ergin SEZGIN}
\footnote{$^\dagger$}
{\tenfoot Supported in part by the U.S.\ National Science Foundation,
under grant \hfill\break
PHY-9106593.}}
\vskip .5truecm
\centerline{Center for Theoretical Physics}
\centerline{Texas A\&M University}
\centerline{College Station, TX 77843-4242, USA}
\vskip 2truecm
\centerline{ABSTRACT}
\vskip .5truecm
Starting from the new minimal multiplet of supergravity in $2+2$
dimensions, we construct two types of self-dual supergravity
theories. One of them involves a self-duality condition on the
Riemann curvature and implies the equations of motion
following from the Hilbert-Einstein type supergravity action.
The other one involves a self-duality condition on a
{\it torsionful} Riemann curvature with the torsion given
by the field-strength of an antisymmetric tensor field,
and implies the equations of motion that follow from an
$R^2$-type action.

\vfill\eject

\noindent{\bf 1. Introduction}

\bigskip

Self-dual supergravity theories in $2+2$ dimensions are interesting to study
for at least two reasons. First, they are candidate backgrounds for the $N=2$
superstring [1]. Second, their suitable reductions to 1+1 dimension
are expected to yield a large class of integrable systems [2]. Recently, a
self-dual supergravity in
$2+2$ dimensions has been constructed [3]. It is essentially obtained
by imposing self-duality and chirality conditions on the curvatures
of an Hilbert-Einstein type supergravity theory.

In this letter we shall construct a new type of self-dual supergravity
in $2+2$ dimensions.
Our construction is
based upon the new minimal multiplet of supergravity [4]
in which the antisymmetric
tensor component of the off-shell supergravity multiplet occurs
as the torsion part of the spin connection field [5]. In this theory
one arrives naturally at a self-duality condition on the
{\it torsionful} Riemann tensor.
We will show that in this type of self-dual supergravity theory
the field equations of
an $R^2$-type action must be satisfied.

It has been shown in [5,6] that supergravity in the
new minimal formulation can be put into a form which is very
similar to the super Yang-Mills system with the Yang-Mills field
replaced by a Lorentz connection with torsion. We shall therefore begin
with a discussion of
self-dual super Yang-Mills theory [3].
We shall show that this theory actually has a hidden superconformal
symmetry which is based on the superalgebra $SL(4|1)$.
We shall then
describe our construction of the two types of self-dual supergravity
theories in $2+2$ dimensions. In the conclusion we will
discuss a number of questions raised by this work.
In the Appendix we give details of
the superconformal algebra $SL(4|1)$ in $2+2$ dimensions.

\bigskip
\noindent{\bf 2. Self-dual Superconformal Yang-Mills Theory}

\bigskip

We start with the description of the super-Poincar\'e algebra in $2+2$
dimensions.
The generators are the supercharges $Q_\pm$, which are pseudo-Majorana-Weyl
spinors
in $2+2$ dimensions [7],
and the Poincar\'e generators, $P_\mu$ and $M_{\mu\nu}$.
The anticommutators between the supercharges are given by
\footnote{$^*$}{\tenfoot
Our conventions are:
$\{\gamma_\mu,\gamma_\nu\}=2\eta_{\mu\nu}$ where
$\eta_{\mu\nu}= {\rm diag}(-,-,+,+)$ and
$\gamma_\mu^*=\gamma_\mu$;\ $C=\gamma_{12}$,\ $C^T=-C$,\
$(\gamma^\mu C)^T=(\gamma^\mu C)$,\ $(\gamma^{\mu\nu}C)^T=\gamma^{\mu\nu}C$,\
$\epsilon_{1234}=+1$;\ $\gamma_5=\gamma_{1234}$,\  $\gamma_5^2=1$,\
$[C,\gamma_5]=0$.
$Q_\pm$ are real two-component independent pseudo-Majorana-Weyl spinors.
The Fierz rearrangement formula is: $\chi\bar\psi=-\ft12 \bar\psi\chi
+\ft18\bar\psi\gamma^{\mu\nu}\chi\gamma_{\mu\nu}$ for same chirality
anticommuting pseudo-Majorana-Weyl spinors and $\chi\bar\psi=-\ft12
\bar\psi\gamma^\mu\chi\gamma_\mu$ for the case of opposite chiralities.
A useful identity is:
$\gamma^{\mu\nu}=-\ft12\epsilon^{\mu\nu\rho\sigma}\gamma_{\rho\sigma}\gamma_5$.
}

$$
\eqalign{
\{Q_+,Q_+\} &= 0 \cr
\{Q_+,Q_-\} &= \ft12 (1+\gamma_5)\gamma_\mu C P^\mu\cr
\{Q_-,Q_-\} &= 0 \cr
}\eqno(1)
$$
{}From the well-known super Yang-Mills theory in $3+1$, it is easy
to extract the analogous result for $2+2$ dimensions, which reads

$$
\eqalignno{
\delta A_\mu &= \ft12 \bar\epsilon_+\gamma_\mu\lambda_- +
\ft12\bar\epsilon_-
\gamma_\mu\lambda_+ &(2)\cr
\delta\lambda_\pm &= - \ft14 \gamma^{\mu\nu}
F_{\mu\nu}\epsilon_\pm \pm \ft12 D\epsilon_\pm &(3)\cr
\delta D &= \ft12 \bar\epsilon_+ \gamma^\mu D_\mu \lambda_-
- \ft12 \bar\epsilon_-\gamma^\mu D_\mu \lambda_+&(4)\cr
}
$$
where $\lambda_\pm$ are pseudo-Majorana-Weyl spinors, $D$ is the real
auxiliary scalar field and $D_\mu$ is the gauge-covariant derivative.
All fields carry an adjoint Yang-Mills index which we have suppressed.
Since this is an off-shell multiplet, no field equations,
including the self-duality condition, are implied by the above
transformation rules. Following [3] we now impose the condition

$$
\lambda_+=0 \eqno(5)
$$
The consequences of this condition are as follows. First
from (3) we obtain $D=0$ and the self-dual Yang-Mills equation

$$
\gamma^{\mu\nu}\epsilon_+ F_{\mu\nu} =0 \quad \Leftrightarrow \quad
F_{\mu\nu} = \ft12 \epsilon_{\mu\nu\rho\sigma} F^{\rho\sigma}\eqno(6)
$$
Next, from (4) we deduce that the nonvanishing spinor $\lambda_-$
satisfies the field equation

$$
\gamma^\mu D_\mu \lambda_- =0\eqno(7)
$$
In summary, the self-dual super Yang-Mills system simply consists of the
field equations (6) and (7) which transform into each other under the
supersymmetry transformations

$$
\eqalignno{
\delta A_\mu &= \ft12 \bar \epsilon_+\gamma_\mu \lambda_- &(8)\cr
\delta \lambda_- &= - \ft14 \gamma^{\mu\nu}F_{\mu\nu}
\epsilon_- &(9)\cr
}
$$
One can check that the closure of the algebra indeed requires
(6) and (7). Note
that both supersymmetries $Q_\pm$ are present
in this system and therefore we have $(1,1)$ type supersymmetry
\footnote{$^*$}{\tenfoot
The truncation to $(1,0)$ supersymmetry by setting $\epsilon_+=0$
would lead to the
superalgebra $\{Q_-,Q_-\}=0$ which is realized trivially
by the transformation rules $\delta A_\mu=0$ and $\delta \lambda_-
= -\ft14 \gamma^{\mu\nu}F_{\mu\nu}\epsilon_-$. The truncation to
$(0,1)$ supersymmetry which is achieved by setting $\epsilon_-=0$
would lead to the superalgebra $\{Q_+,Q_+\}=0$, again realized
trivially by the transformation rules
$\delta\lambda_-=0$ and $\delta A_\mu = \ft12\bar \epsilon_+
\gamma_\mu\lambda_-$.
}.
Notice that despite supersymmetry, the self-duality equation (6) receives no
fermionic modifications.

It is well-known that the self-dual Yang-Mills equations are invariant
under the conformal group of transformations, which is the group $SO(3,3)$
for a spacetime with signature $(2,2)$. Not surprisingly, the self-dual
super Yang-Mills equations (6) and (7) are invariant under the  superconformal
group $SL(4|1)$ whose bosonic subgroup is $SL(4) \approx SO(3,3)$.
These transformations are given by (8), (9) and

$$
\eqalignno{
\delta A_\mu &= -\ft12 \bar\eta_-\gamma^\nu\gamma_\mu\lambda_-
x_\nu + \xi^\lambda\partial_\lambda A_\mu &(10)\cr
\delta\lambda_- &= -\ft14 \gamma^{\mu\nu}\gamma^\rho x_\rho F_{\mu\nu}
\eta_+ + \ft32\alpha\lambda_- - \beta\lambda_-
+\ft14\omega^{\mu\nu}\gamma_{\mu\nu}\lambda_-
+\xi^\mu\partial_\mu\lambda_- &(11)\cr
}
$$
where $\eta_\pm$ are the special conformal supersymmetry parameters,
$\alpha,\beta$ are the parameters of dilatations and $SO(1,1)$
transformations, respectively, and $\xi^\mu$ represent the
spacetime conformal transformations given by

$$
\xi^\mu(x)= a^\mu - \omega^{\mu\nu}x_\nu + \alpha x^\mu
+2x^\mu\eta \cdot x - \eta^\mu x^2
\eqno(12)
$$
Here $a^\mu,\omega^{\mu\nu}$ and $\eta^\mu$ are the constant
parameters of translations, Lorentz rotations and conformal boosts,
respectively. The above superconformal transformations indeed satisfy
the superconformal algebra $SL(4|1)$ given in the Appendix.

\bigskip
\noindent{\bf 3. Self-dual Supergravity Theory}
\bigskip
In $3+1$ dimensions there are many different versions of off-shell
supergravity. Similar off-shell supergravities in $2+2$
dimensions can be easily obtained from them. We shall work with the
new minimal formulation. As has been shown in [5,6] this supergravity
theory can be put into a form which is very similar to the
super Yang-Mills system with the Yang-Mills field replaced by
a Lorentz connection with torsion. This is a convenient
feature since it allows us to apply the procedure of the previous
section to obtain self-dual supergravity equations.

The new mimimal multiplet [4] contains the vierbein $e_\mu{}^a$, the
gravitini $\psi_{\mu\pm}$, the antisymmetric tensor field
$B_{\mu\nu}$ and an $SO(1,1)$ gauge field $V_\mu$. In $2+2$
dimensions the supersymmetry transformations are

$$
\eqalignno{
\delta e_\mu{}^a &= \ft12 \bar\epsilon_+\gamma^a\psi_{\mu-}+
\ft12\bar\epsilon_-\gamma^a\psi_{\mu+} &(13)\cr
\delta \psi_{\mu\pm} &= {\cal D}_\mu(\Omega_+,V_+)\epsilon_\pm &(14)\cr
\delta B_{\mu\nu} &= \ft32\bar\epsilon_+\gamma_{[\mu}\psi_{\nu]-}
 + \ft32\bar\epsilon_-\gamma_{[\mu}\psi_{\nu]+} &(15)\cr
\delta V_\mu &= \ft18 \bar\epsilon_+\gamma_\mu \gamma^{ab}
\psi_{ab-} - \ft18\bar\epsilon_-\gamma_\mu\gamma^{ab}
\psi_{ab+} &(16)
\cr}
$$
where we have defined the combinations\footnote{$^*$}{\tenfoot
The supercovariant spin-connection $\omega_\mu{}^{ab}(e,\psi)$ is
defined as follows:\hfill\eject
$$\omega_\mu{}^{ab}(e,\psi) = \omega_\mu{}^{ab}(e)
- \ft12 \bar\psi_\mu\gamma^{[a}\psi^{b]} -\ft14\bar\psi^a\gamma_\mu\psi^b
$$
where $\psi_\mu=\psi_{\mu+}+\psi_{\mu-}$.
}

$$
\eqalignno{
\Omega_{\mu\pm}{}^{ab} &= \omega_\mu{}^{ab}(e,\psi) \pm
\hat H_{\mu }{}^{ab} &(17)\cr
V_{\mu+} &= V_\mu + \ft16\epsilon_\mu{}^{abc}\hat H_{abc} &(18)
\cr}
$$
The covariant curvatures are

$$
\eqalignno{
\psi_{\mu\nu\pm} &= {\cal D}_\mu(\Omega_+,V_+)\psi_{\nu\pm}
- {\cal D}_\nu(\Omega_+,V_+)\psi_{\mu\pm}
&(19)\cr
\hat H_{\mu\nu\rho} &=\partial_{[\mu} B_{\nu\rho]} -
\ft32\bar\psi_{[\mu+}\gamma_\nu\psi_{\rho]-} &(20)
}
$$
The derivative ${\cal D}_\mu$ on the supersymmetry parameters $\epsilon_\pm$
is given by

$$
{\cal D}_\mu(\Omega_+,V_+)\epsilon_\pm = \biggl (
\partial_\mu - \ft14 \Omega_{\mu+}{}^{ab}\gamma_{ab} \mp
V_{\mu+}\biggr ) \epsilon_\pm \eqno(21)
$$
Following [5,6], one can show that $\Omega_{\mu -}{}^{ab}, \psi^{ab}$
and the supercovariant field-strength \hfill\break $\hat F^{ab}(V_+)$
form a
super-Yang-Mills type multiplet with the gauge group taken to be
the spacetime Lorentz group $SO(2,2)$.
We find that the transformation rules are\footnote{$^\dagger$}{
\tenfoot The
supercovariant curvatures of $V_{\mu+}$ and $\Omega_{\mu-}{}^{ab}$
are defined as follows: \hfill\eject

$$\hat F_{ab}(V_+)=2\partial_{[a}V_{b]+}-\ft12 \bar\psi_{[a+}\gamma^\lambda
\psi_{b]\lambda-} + \ft12 \bar\psi_{[a-}\gamma^\lambda\psi_{b]\lambda+}$$
and
$$\hat R_{cd}{}^{ab}(\Omega_-) = R_{cd}{}^{ab}(\Omega_-)
-\bar\psi_{[c+}\gamma_{d]}\psi_-^{cd} -
\bar\psi_{[c-}\gamma_{d]}\psi_+^{cd}$$\ \ \ \ .
}

$$
\eqalignno{
\delta \Omega_{\mu -}{}^{ab} &=
\ft12\bar\epsilon_+\gamma_\mu \psi^{ab}_- +
\ft12\bar\epsilon_-\gamma_\mu \psi^{ab}_+ &(22)\cr
\delta \psi^{ab}_\pm &= - \ft14 \gamma^{cd}\hat R_{cd}{}^{ab}(\Omega_-)
\epsilon_\pm \mp \hat F^{ab}(V_+)\epsilon_\pm &(23)\cr
\delta \hat F^{ab}(V_+) &=
-\ft14\bar\epsilon_+\gamma^\mu
D_\mu(\Omega_-)\psi^{ab}_- +
\ft14\bar\epsilon_-\gamma^\mu
D_\mu(\Omega_-)\psi^{ab}_+ &(24)\cr
}
$$
where $D_\mu(\Omega_-)$ is the supercovariant derivative and
the $\Omega_-$ covariantization in the last line acts on the
fermionic as well as the vectorial indices of $\psi^{ab}_\pm$.

Below we shall construct two types of self-dual supergravity theories
which we shall refer to as type I and type II. To obtain the type
I self-dual supergravity we shall exploit the fact that the
transformation rules (22)-(24) have the super Yang-Mills form and use the
procedure of the previous section.
This leads to the self-dual supergravity theory of [3].
To obtain the type II self-dual
supergravity we shall make use of the fact the Yang-Mills group
in (22)-(24) is actually the spacetime group $SO(2,2)$.
\bigskip

\centerline{\it Type I Reduction}
\bigskip
Following the procedure of the previous section
we impose the condition

$$
\psi_+^{ab}=0 \eqno(25)
$$
The consequence of this can be seen from (23) and (24) to be

$$
\eqalignno{
\hat R_{cd}{}^{ab}(\Omega_-) &= \ft12 \epsilon_{cdef}\hat R^{efab}
(\Omega_-)  &(26)\cr
\hat F^{ab}(V_+) &=0 &(27)\cr
\gamma^\mu D_\mu(\Omega_-)\psi^{ab}_- &= 0 &(28)\cr
}
$$
We recognize (27) as one of the equations of motion
that follow from the $2+2$ version
of the $R$-type action of [4]

$$
e^{-1}{\cal L}(R) = -\ft12R(\omega) - \bar\psi_{\mu+}
\gamma^{\mu\nu\rho}{\cal D}_\nu(\omega,V)\psi_{\rho-}
-\ft12\hat H_{abc}\hat H^{abc} + \ft23 e^{-1}\epsilon^{\mu\nu\lambda
\rho}V_\mu\partial_\nu B_{\lambda\rho} \eqno(29)
$$
By supersymmetry, the remaining equations of motion based
on this action also follow. As for (28), it can be written in terms
of derivatives of the gravitino and Einstein equations of motion.

One of the equations of motion that follows from (29) is

$$
\hat H_{abc}=0\eqno(30)
$$
Hence, (26) reduces to

$$
R_{cd}{}^{ab}(\omega) = \ft12 \epsilon_{cdef} R^{efab}
(\omega)  \eqno(31)
$$
where we have used (25). From (31) and the identity
$R_{[abc]d}(\omega)=0$, it then follows that the Ricci-tensor vanishes, i.e.

$$
R_{ab}(\omega)=0 \eqno(32)
$$
The supersymmetry variation of (31) leads to the
following duality condition on the gravitino curvature:

$$
\psi^{ab}_- =  \ft12\epsilon^{abcd}\psi_{cd-}\eqno(33)
$$
This duality condition can also directly be proven from the
gravitino field equation $\gamma_a\psi^{ab}_-=0$.

In summary, the type I self-dual supergravity [3] is described by (31) and (33)
plus the Bianchi identities for the curvatures which together imply
the equations of motion that follow from the $R$-action.
Due to (25), (27) and (30) the fields
$\psi_{\mu+}$, $V_\mu$ and $B_{\mu\nu}$
drop out of the theory and the only equations of motions
in addition to (31) and (33) are the
supercovariant Einstein and gravitino field equations for the
fields $e_\mu{}^a$ and $\psi_{\mu-}$, respectively.
Thus, the type I self-dual supergravity has the local supersymmetry
given by

$$\eqalignno{
\delta e_\mu{}^a &= \ft12\bar\epsilon_+\gamma^a\psi_{\mu-}&(34)\cr
\delta \psi_{\mu-} &= {\cal D}_\mu(\omega)\epsilon_- &(35)\cr
}
$$
Closure of the algebra requires that the supersymmetry parameter
$\epsilon_+$ be covariantly constant, i.e.

$$
D_\mu(\omega)\epsilon_+=0\eqno(36)
$$
The integrability condition of this equation leads to the self-duality
condition (31) for the Riemann curvature.

Note that despite that fact that there are no fermionic corrections to the
self-duality equation (31) it nonetheless is supercovariant and transforms
into the gravitino field equation under the above
supersymmetry transformations.
Note also that, since the self-duality conditions (31) and (33)
do not follow from
the action (29), of course the latter can not be considered as the action
for type I self-dual supergravity.

\bigskip

\centerline{\it Type II Reduction}
\bigskip
In this reduction we make use of the
fact that each one of the gravitino curvatures  $\psi^{ab}_\pm$ is
in a reducible representation of $SO(2,2)$, namely $(3,1)\oplus(1,3)$.
The type II reduction amounts to setting one of these pieces equal
to zero, i.e.

$$
\psi^{ab}_\pm =  \ft12\epsilon^{abcd} \psi_{cd\pm}  \eqno(37)
$$
This condition and supersymmetry are evidently consistent with
the following self-duality conditions

$$
\eqalignno{
\hat R_{cd}{}^{ab}(\Omega_-) &= \ft12 \epsilon^{abef}\hat R_{cdef}
(\Omega_-)
&(38)\cr
\hat F^{ab}(V_+) &= \ft12\epsilon^{abcd}\hat F_{cd}(V_+) &(39)\cr
}
$$
The above three equations, together with the Bianchi identities for
the curvatures, define type II self-dual supergravity.
We observe that (39) together with the Bianchi identity

$$
D_{[a}\hat F_{bc]}(V_+)
+ \ft14\bar\psi_{[ab+}\gamma^\lambda\psi_{c]\lambda-}
- \ft14\bar\psi_{[ab-}\gamma^\lambda\psi_{c]\lambda+}
=0\eqno(40)
$$
and the self-duality conditions (37) imply that

$$
8D_a \hat F^{ab}(V_+) + \bar\psi^{cd}_+\gamma^b\psi^{cd}_- = 0
\eqno(41)
$$
We recognize this as one of the field equations that
follow from
the $2+2$ version of the $R^2$-type supergravity action
constructed in [5,6]

$$\eqalign{
e^{-1}{\cal L}(R^2) =& \ft14 R_{\mu\nu}{}^{ab}(\Omega_-)
R^{\mu\nu ab}(\Omega_-) - 2 \hat F_{ab}(V_+)\hat F^{ab}(V_+)
+ \ft12\bar\psi^{ab}\gamma^\mu D_\mu(\Omega_-,V)
\psi_{ab} \cr
&+\ft18\bar\psi_\mu\gamma^{cd}\gamma^\mu\psi_{ab} \bigl (
R_{cd}{}^{ab}(\Omega_-)+\hat R_{cd}{}^{ab}(\Omega_-)\bigr )
\cr}\eqno(42)
$$
where $\psi_\mu=\psi_{\mu+}+\psi_{\mu-}$ and the $\Omega_-$ covariantisation
in the kinetic term of the gravitino curvature acts both on the spinor
as well as the vector indices of $\psi^{ab}$.
By supersymmetry it then follows that the field equations for the remaining
fields $e_\mu{}^a, \psi_{\mu\pm}$ and $B_{\mu\nu}$ must also be satisfied.
We arrive at the same conclusion by considering the consequences of (37)
and (38) together with the Bianchi identities.

In summary, the type II self-dual supergravity theory consists of
the self-duality equations (37)-(39)
plus the Bianchi identities for the curvatures which together imply
the equations of motion that follow from the $R^2$-action. Note that
in contradistinction to the type I theory the selfduality condition
on the Riemann curvature involves extra fermionic terms. The type II
theory is invariant under the following supersymmetry transformations

$$
\eqalignno{
\delta e_\mu{}^a &= \ft12 \bar\epsilon_+\gamma^a\psi_{\mu-}+
\ft12\bar\epsilon_-\gamma^a\psi_{\mu+} &(43)\cr
\delta\psi_{\mu\pm} &= (\partial_\mu -\ft14\Omega_{\mu+}{}^{ab}\gamma_{ab}
\mp V_{\mu+})\epsilon_\pm &(44)\cr
\delta B_{\mu\nu} &= \ft32\bar\epsilon_+\gamma_{[\mu}\psi_{\nu]-}
 + \ft32\bar\epsilon_-\gamma_{[\mu}\psi_{\nu]+} &(45)\cr
\delta V_\mu &= \ft18\bar\epsilon_+\gamma_\mu\gamma^{ab}
\psi_{ab-} &(46)
\cr}
$$

Although the equations of motion of type II
self-dual supergravity imply the equations of motion corresponding
to the above $R^2$ action, the converse is not true and therefore
the $R^2$ action, of course, can not be considered as the action for
the type II self-dual supergravity.

Finally, we note that the self-duality condition (38) implies that

$$
\hat R_{cd}{}^{ab}(\Omega_+) = \ft12 \epsilon_{cdef}\hat R^{efab}
(\Omega_+)  \eqno(47)
$$
owing to the identity

$$
R_{abcd}(\Omega_-)=R_{cdab}(\Omega_+)\eqno(48)
$$
Furthermore, for a {\it torsionful} curvature the self-duality condition
(38) does {\it not} imply that the
Ricci tensor vanishes. Instead, one is only able to show that the Ricci
tensor is proportional to torsion-dependent terms.

\bigskip

\noindent{\bf 4. Discussion}

\bigskip

In this letter we have constructed a new type of self-dual supergravity
theory in $2+2$ dimensions based on a {\it torsionful} curvature tensor.
Our results raise a number of interesting
questions.

A natural question to ask is the connection between the self-dual
supergravities considered here and the $N=2$ supersymmetric string
theories. In this regard we recall that {\it off-shell} supergravity
in the new minimal formulation is a consistent background for the
Green-Schwarz superstring in $3+1$ dimensions [8].
The same model can be taken over to represent the Green-Schwarz
superstring in $2+2$ dimensions [3], with obvious changes in the
signature of the spacetime metric and the reality conditions on the
spinorial coordinates of the target superspace, which will now be
parametrized by $x^\mu, \theta_+$ and $\theta_-$. An interesting open
problem is to find a mechanism that will put the supergravity background
on-shell, with or without self-duality conditions.

It is interesting to note that the mathematics used in discussing self-dual
Yang-Mills or supergravity theories resembles that used in the discussion
of Kaluza-Klein compactifications and solitonic solutions
to string theories in ten dimensions. An
important difference, however, is that in the latter case all
background fermions are set equal to zero and not all supersymmetries
are maintained, depending on the background. Typically, residual
supersymmetries are found in self-dual backgrounds. In the case
of self-dual supergravities all supersymmetries are maintained
and not all fermions are vanishing.

Finally, we recall that the supergravities considered in this letter have
a $(1,1)$ supersymmetry in $2+2$ dimensions. A chiral truncation
to $(1,0)$ supersymmetry would lead to the superalgebra $\{Q_-,Q_-\}
=0$. In this case there is no need to supersymmetrize the
bosonic self-duality conditions since the gauge fields will be inert
under supersymmetry. To obtain a nontrivial $(1,0)$-supersymmetric
self-dual systems in $2+2$ dimensions, one should consider the
following superalgebra:

$$
\{Q_-,Q_-\} = \ft12\gamma^{\mu\nu}CM_{\mu\nu}
\eqno(49)
$$
where $M_{\mu\nu}$ are the Lorentz generators, self-dual in the $\mu,\nu$
indices, therefore representing effectively an $SO(2,1)$ rotation.
A $(1,0)$ self-dual supergravity theory based upon this algebra
has been considered in [9].
\bigskip
\bigskip
\centerline{\bf ACKNOWLEDGEMENTS}
\bigskip

We are grateful to Mees de Roo for useful discussions and we thank
Chris Pope for a careful reading of the manuscript.
E.S.\ would like to thank the institute for Theoretical Physics
in Groningen where this work was done for its hospitality.
For one of us (E.B.) this work has been made possible by a fellowship
of the Royal Netherlands Academy of Arts and Sciences (KNAW).

\vfill\eject

\centerline{\bf APPENDIX: The Superconformal Algebra $SL(4|1)$}

\bigskip

The bosonic subalgebra of $SL(4|1)$ is $SL(4) \times SO(1,1)$
generated by the $SO(2,2)$ generators $M_{\mu\nu}$, translations
$P_\mu$, conformal boosts $K_\mu$, dilatation $D$ and the $SO(1,1)$
generator $A$. The nonvanishing
commutators for this bosonic subalgebra are

$$
\eqalign{
[M_{\mu\nu}, M_{\rho\sigma}] &= \eta_{\nu\rho}M_{\mu\sigma}
-\eta_{\nu\sigma}M_{\mu\rho} - \eta_{\mu\rho}M_{\nu\sigma} +
\eta_{\mu\sigma}M_{\nu\rho}\cr
[M_{\mu\nu}, P_\rho]&=\eta_{\nu\rho}P_\mu - \eta_{\mu\rho}P_\nu
\quad\quad
[M_{\mu\nu}, K_\rho]=\eta_{\nu\rho}K_\mu
-\eta_{\mu\rho}K_{\nu} \cr
[P_\mu, K_\nu]&=2(\eta_{\mu\nu}D-M_{\mu\nu}) \quad\quad
[P_\mu,D]=P_\mu \quad\quad
[K_\mu,D]=-K_\mu \cr}
$$

The fermionic generators are the supersymmetry generators $Q_\pm$ and the
special supersymmetry generators $S_{\pm}$, all of which are real two
component pseudo Majorana-Weyl spinors. The nonvanishing (anti)commutators
involving the spinorial generators are
$$
\eqalign{
\{Q_-, Q_+\} &=-2(P_-\gamma^\mu C)P_\mu \quad\quad
\{S_+,S_-\}  = 2(P_+\gamma^\mu C)K_\mu \cr
\{S_\pm,Q_\pm \}&=P_\pm (2D+\gamma^{\mu\nu}M_{\mu\nu}\pm A)C \cr
[P_\mu, S_\pm ]&=-\gamma_\mu Q_\mp  \quad\quad
[K_\mu, Q_\pm] = \gamma_\mu S_\mp \cr
[M_{\mu\nu}, Q_\pm]&=-\ft12 \gamma_{\mu\nu}Q_\pm \quad\quad
[M_{\mu\nu}, S_\pm]=-\ft12 \gamma_{\mu\nu} S_\pm \cr
[D, Q_\pm] &=-\ft12 Q_\pm \quad\quad
[D, S_\pm] =\ft12 S_\pm \cr
[A, Q_\pm] &=\pm 3Q_\pm \quad\quad
[A, S_\pm] = \mp 3S_\pm \cr}
$$
where $P_\pm = \ft12 (1\pm\gamma_5)$.

\vfill\eject

\centerline{\bf REFERENCES}

\bigskip

\item{1.} H.\ Ooguri and C.\ Vafa, Mod.\ Phys.\ Lett.\ {\bf A5} (1990)
1389; Nucl.\ Phys.\ {\bf B361} (1991) 469; Nucl.\ Phys.\ {\bf B367} (1991)
83.
\item{2.} See, for example, the review article by R.S.\ Ward in
{\it Field Theory, Quantum Gravity and Strings}, Vol.\ 2, p.\ 106,
eds.\ H.J.\ De Vega and N.\ Sanchez.
\item{3.} S.J.\ Gates Jr., H.\ Nishino and S.V.\ Ketov,
{\it Extended Supersymmetry and Self-Duality in $2+2$ Dimensions},
preprint, UMDEPP 92-187;
{\it Supersymmetric Self-Dual Yang-Mills and Supergravity as Background of
Green-Schwarz Superstring},
preprint, UMDEPP 92-171;
{\it Majorana-Weyl Spinors and Self-Dual Gauge Fields in
$2+2$ Dimensions}, preprint, UMDEPP 92-163.
\item{4.} M.F.\ Sohnius and P.C.\ West, Phys.\ Lett.\ {\bf 105B}
(1981) 353.
\item{5.}
S.\ Ferrara, S.\ Sabharwal and
M.\ Villasante, Phys.\ Lett.\ {\bf 205B} (1988) 302;
S.\ Cecotti, S.\ Ferrara, M.\ Porrati and S.\ Sabharwal,
Nucl.\ Phys.\ {\bf B306} (1988) 160.
\item{6.} M.\ de Roo, A.\ Wiedemann and E.\ Zijlstra,
Class.\ Quantum Grav.\ {\bf 7} (1990) 1181.
\item{7.} T.\ Kugo and P.K.\ Townsend, Nucl.\ Phys.\ {\bf B221} (1983)
357.
\item{8.} E.\ Bergshoeff, E.\ Sezgin and P.K.\ Townsend,
Phys.\ Lett.\ {\bf 169B} (1986) 191.
\item{9.} L.J.\ Romans, unpublished.

\end